\documentclass[a4paper,aps,amsmath,latexsym,amsfonts]{jpconf}

\usepackage{graphicx}
\usepackage{amsfonts,amssymb,amsmath}

\def\half{\textstyle{\frac{1}{2}}}
\def\quarter{\textstyle{\frac{1}{4}}}

\newcommand{\be}{\begin{equation}}
\newcommand{\ee}{\end{equation}}
\newcommand{\bea}{\begin{eqnarray}}
\newcommand{\eea}{\end{eqnarray}}

\begin{document}
\rightline{DCPT-10/67}
\title{Holographic Superconductivity with Gauss-Bonnet gravity}

\author{Ruth Gregory}

\address{Centre for Particle Theory
Durham University, South Road, Durham, DH1 3LE, UK}

\begin{abstract}
I review recent work (\cite{GKS,BGKS}) on holographic superconductivity
with Einstein-Gauss-Bonnet gravity, and show how the
critical temperature of the superconductor depends on both
gravitational backreaction and the Gauss-Bonnet parameter, using
both analytic and numerical arguments.
I also review computations of the conductivity,
finding the energy gap, and demonstrating that there is no
universal gap ratio, $\omega_g/T_c$, for these superconductors.
\end{abstract}

\section{Introduction}

Holographic superconductivity (see \cite{Herzog,Hartnoll,Gary1} for reviews)
is a fascinating idea which proposes to describe strongly coupled 
high $T_c$ superconductors via a classical gravitational system using the 
gauge gravity correspondence, \cite{Malda}. 
The bulk theory typically has gauge and a charged scalar field,
with a black hole providing a finite temperature.
Choosing an appropriate scalar potential, such as a negative
mass-squared above the Breitenlohner-Friedman bound, \cite{BF},
allows the scalar to condense out of the vacuum near the
black hole horizon, if the curvature is sufficiently large
there \cite{Gub}. This occurs for low temperature black holes, hence the
analogy to superconductivity.

The superconducting phase corresponds to a ``hairy'' black hole,
where the condensation of the scalar field out of its symmetric
state screens the charge of the black hole. Typical no-hair theorems
\cite{nohair} do not apply in this case as the scalar potential is not
positive definite.
Near the boundary, the scalar has a power law fall-off, and the coefficient
of this fall-off can be interpreted as a condensate in the 
boundary theory. Following the behaviour of this condensate as
the temperature drops reveals the typical behaviour for an order
parameter governing the superconducting phase transition, and has
been observed in a wide variety of bulk thoeries, including varying 
the scalar mass and potential, varying the 
number of spacetime dimensions, as well as
having magnetic fields present and the stability of the 
system \cite{Gub,HHH,HorRob,zeroT,DIM,magnetic,KS}.
Similar models have also been found from embeddings in string
theory, \cite{STSC}, and it is therefore of interest to 
consider more general stringy aspects of these models.

In this presentation, I review the results of \cite{GKS} and \cite{BGKS}, 
which explored adding the higher curvature Gauss-Bonnet (GB) 
term, \cite{GBL}, to the gravitational action.
I will also update on the talk where appropriate, and add a few
new results and comments.
In brief, many of the qualitative 
features of the holographic superconductors are stable under 
higher curvature corrections, however, there are quantitative
differences, in particular, a universality relation for an energy
gap in the superconductor was found not to be substantiated 
even in the Einstein limit.
(See also \cite{GBscs} for other work on GB holographic superconductors.)

I first review the bulk superconductor, deriving some general
analytical results and remarking on the probe superconductor. Then
I summarize the numerical results, showing how the numerical results 
follow closely an analytical bound, revealing an interesting subtlety in the
response of the superconductor to the Gauss-Bonnet parameter $\alpha$.
Finally, I review the results of \cite{GKS} and \cite{BGKS} on
conductivity.

\section{The bulk superconductor}
\label{sec:SC}

The general action is that of an Einstein-Gauss-Bonnet (EGB) 
gravitational action coupled
to a massive charged complex scalar field and a U(1) gauge field:
\begin{eqnarray}
S&=&\frac{1}{2\kappa^2}\int d^5x \sqrt{-g} \left[
-R + \frac{12}{L^2} + \frac{\alpha}{2} \left(
R^{abcd}R_{abcd} -4R^{ab}R_{ab} + R^2
\right) \right]\nonumber\\
&&\hspace{3.5mm}+\int d^5x\sqrt{-g}\left[
-\frac{1}{4}F^{ab}F_{ab}+|\nabla_a\psi -iqA_a\psi|^2
-m^2|\psi|^2
\right]\,
\label{action}
\end{eqnarray}
where $g$ is the determinant of the metric, and $R_{abcd}$, $R_{ab}$ 
and $R$ are the Riemann curvature tensor, Ricci tensor, and the Ricci 
scalar, respectively.  The Gauss-Bonnet coupling constant is taken in 
the range $\alpha\in[0,L^2/4]$, and
the cosmological constant is given in terms of a length scale, $L$, which 
for Einstein gravity ($\alpha=0$) corresponds to the adS curvature. 
The Planck scale is set by $\kappa^2 = 8\pi G_5$, $q$ is the charge, 
and $m^2$ the (squared) mass, of the scalar field.

The equations of motion satisfied by the bulk system are:
\be
\label{GBeqns}
R_{ab} - {\half} R g_{ab} + \frac{6}{L^2} g_{ab}
-\alpha \left [ H_{ab} - \quarter H g_{ab} \right ] = \kappa^2 T_{ab}
\ee
where
\be
\label{Hdef}
H_{ab} = R_a^{\;cde} R_{bcde} - 2 R_{ac} R^c_b - 2 R_{acbd} R^{cd}
+R R_{ab} \;,
\ee
$T_{ab}$ is the matter energy momentum tensor 
\be
\label{tabdef}
T_{ab} = 2 {\cal D}_{(a} \psi^\dagger {\cal D}_{b)} \psi
-F_{ac} F_b^{\;c} - \left [
|{\cal D}_c \psi |^2 - \quarter F_{cd}^2 - m^2 |\psi|^2
\right ]g_{ab} \;,
\ee
and ${\cal D}_a = \nabla_a - iqA_a$ is the gauge covariant derivative.

To examine holographic superconductivity, we allow for a bulk black hole
in order to have a system at finite temperature:
\begin{eqnarray}
ds^2 = f(r) e^{2\nu(r)} dt^2 - \frac{dr^2}{f(r)} 
- \frac{r^2}{L_e^2}(dx^2+dy^2+dz^2)
\end{eqnarray}
where
\be
f(r) = \frac{r^2}{2\alpha} \left [ 1 - \sqrt{
1 - \frac{4\alpha}{L^2} \left ( 1 - \frac{r_+^4}{r^4} \right )}
\right]
\ee
is the black hole gravitational potential in the absence of backreaction, and
\be
\label{Leff}
L^2_{\rm e}=\frac{L^2}{2} \left[1+\sqrt{1-\frac{4\alpha}{L^2}}\right]
\ee
is the actual curvature of the adS spacetime, which 
is renormalized away from the cosmological constant scale, $L$, 
once $\alpha$ is nonzero. The temperature of the background is:
\begin{eqnarray}
T = \frac{1}{4\pi} f' (r_+)e^{\nu(r_+)}\ ,
\label{hawking}
\end{eqnarray}
which is equal to $r_+/\pi L^2$ in the absence of backreaction.

Taking a static ansatz, $A_\mu=(\phi(r),0...)$ and $\psi =\psi(r)$
(which can be taken to be real), the equations of motion become
\begin{eqnarray}
&&\phi^{\prime\prime}+\left( \frac{3}{r}-\nu^\prime
\right)\phi^\prime -2q^2\frac{\psi^2}{f}\phi=0\,, \label{phieq}\\
&&\psi^{\prime\prime}+\left( \frac{3}{r}+\nu^\prime+\frac{f^\prime}{f}
\right)\psi^\prime +\left(\frac{q^2\phi^2}{f^2e^{2\nu}}
-\frac{m^2}{f} \right)\psi=0\,, \label{psieq}\\
&&\left(1-\frac{2\alpha f}{r^2} \right)\nu^\prime
=\frac{2\kappa^2}{3}r\left(
\psi^{\prime 2}+\frac{q^2\phi^2\psi^2}{f^2e^{2\nu}}\right) \label{nueq}\\
&&\left(1-\frac{2\alpha f}{r^2} \right)f^\prime+\frac{2}{r}f
-\frac{4r}{L^2} =-\frac{2\kappa^2}{3}r\left[
\frac{\phi^{\prime2}}{2e^{2\nu}}+m^2\psi^2+
f\psi^{\prime2}+\frac{q^2\phi^2\psi^2}{fe^{2\nu}} \right] \label{feq}
\end{eqnarray}
where a prime denotes derivative with respect to $r$.
These equations have several scaling symmetries, as 
noted in \cite{BGKS}, which are used to set $L=Q=q=1$.
Note that by fixing $Q=1$, the charge parameter is kept
fixed in all numerical computations.

\begin{enumerate}

\item
$r \to ar$, $t, x^i \to at, ax^i$,
$L \to aL$, $q \to q/a$, $\alpha \to a^2 \alpha$, $A \to a A$.

\item
$r \to br$, $t \to t/b$, $x^i \to x^i/b$,
$f \to b^2f$, $\phi \to b\phi$.

\item
$\phi \to c \phi$, $\psi \to c\psi$, $q \to q/c$, $\kappa^2 \to \kappa^2/c^2$.

\end{enumerate}

The horizon is defined in general by $f(r_+) = 0$, and demanding
regularity of the solution at both the horizon and boundary gives the
following boundary conditions:

\noindent $\bullet$ Horizon:
\begin{eqnarray}
&&
\phi(r_+)=0,\hspace{1cm}
\psi^\prime(r_+)=\frac{m^2}{f^\prime(r_+)}\psi(r_+) \\
&&
\nu^\prime(r_+)=\frac{2\kappa^2}{3}r_+\left(
\psi^\prime(r_+)^2
+\frac{\phi^\prime(r_+)^2\psi(r_+)^2}{f^\prime(r_+)^2e^{2\nu(r_+)}}
\right)\\
&&
f^\prime(r_+)=\frac{4}{L^2}r_+
-\frac{2\kappa^2}{3}r_+\left(
\frac{\phi^\prime(r_+)^2}{2e^{2\nu(r_+)}}
+m^2\psi(r_+)^2
\right)
\end{eqnarray}
\noindent $\bullet$ Boundary:
\begin{eqnarray}
&&\nu \to 0 \;\;\;,\;\;\;\;
f(r) \sim \frac{r^2}{L_e^2}\;\;\; \Rightarrow \\
&& \phi(r) \sim P - \frac{Q}{r^2}\,,\hspace{1cm} 
\psi(r) \sim \frac{C_{-}}{r^{\Delta_-}}+\frac{C_{+}}{r^{\Delta_+}}\,,
\;\;\;\;{\rm as} \;\; r \to \infty\,,
\label{r:boundary}
\end{eqnarray}
where $\Delta_\pm=2\pm\sqrt{4+m^2L_e^2}$ for a general mass $m^2$. 
We choose $C_-=0$, then $P$ and $C_+$ are fixed by consistency
with the near horizon solution.
According to the adS/CFT correspondence, we can interpret 
$ \langle {\cal O}_{\Delta_+} \rangle \equiv C_{+}$, 
where ${\cal O}_{\Delta_+}$ is the operator with the conformal
dimension $\Delta_+$ dual to the scalar field.

In \cite{GKS} we chose to set the mass of the scalar field as $m^2=-3/L^2$, 
so that the mass remained the same as $\alpha$ was varied. However, 
the variation of the effective asymptotic AdS curvature,
(\ref{Leff}), with $\alpha$ relative to $L$ means that this mass
actually {\it increases} (i.e.\ becomes less negative) with respect
to the asymptotic AdS scale, and therefore the dimension of the 
boundary operator corresponding to $C_+$ varies with $\alpha$. Fixing
$m^2 = -3/L_{\rm eff}^2$ relative to the asymptotic AdS scale, 
avoids this problem, however the mass now varies with respect
to the physical mass and temperature of the black hole as $\alpha$
varies. See \cite{Luke} for a detailed examination of the effect
of scalar mass on the holographic superconductor.

\subsection{Analytic bounds}

While the equations of motion (\ref{phieq}-\ref{feq}) require
a numerical solution in general, even in the absence of backreaction,
near the critical temperature we can glean a measure of analytic 
information using the (uncondensed) charged black hole solution, \cite{GBBH}: 
\bea
\label{phibk}
A &=& \phi_0(r) dt 
= \frac{Q}{r_+^2} \left ( 1 - \frac{r_+^2}{r^2} \right) dt \\
\nu_0 &=& 0 \\
f_0(r) &=& \frac{r^2}{2\alpha} \left [ 1 - \sqrt{
1 - \frac{4\alpha}{L^2} \left ( 1 - \frac{r_+^4}{r^4} \right )
+ \frac{8\alpha\kappa^2 Q^2}{3r^4 r_+^2} \left ( 1 - \frac{r_+^2}{r^2}
\right )} \right]
\label{fbk}
\eea
and linearizing in the scalar condensate. In the above,
$Q$ is the charge of the black hole (up to a geometrical factor
of $4\pi$), and $r_+$ is the event horizon, which determines the ``ADM''
mass of the black hole \cite{ADMass}. To avoid a naked
singularity, we restrict the parameter range as $\alpha\leq L^2/4$.

Linearizing in the scalar condensate leaves the background solution for 
the metric and gauge field unchanged to leading order, and we can focus on 
(\ref{psieq}) with $f$ and $\phi$ taking their background values.
Analytic bounds on the critical temperature can be obtained by looking
for simple relations that must hold for the existence of a nontrivial
scalar solution. First consider the
variable $X_n = r^{n}\psi$, which satisfies to leading order: 
\be
X_n'' + \left ( \frac{f_0'}{f_0} - \frac{3-2n}{r} \right ) X_n' + \left (
\frac{q^2\phi^2}{f_0^2} - \frac{m^2}{f_0} - \frac{nf_0'}{rf_0} 
+ \frac{n(n-2)}{r^2} \right ) X_n = 0 \; .
\label{Xeqn}
\ee

To get an upper bound, let $n=2$, and consider the general properties of
a solution.  At the horizon,
\be
X_2'(r_+) = \frac{X_2(r_+)}{4\pi T_c} \left ( \frac{8}{L^2} +m^2 
- \frac{8\kappa^2 Q^2}{3r_+^6} \right ) 
\ee
which is positive for small $\kappa^2 Q^2$ (taking $X_2(r_+)>0$ without
loss of generality). Since $X_2\sim 1/r^{2+\Delta_+}$ as $r\to \infty$, 
the solution must have a maximum for some $r$, which requires that
\be
\left ( \frac{q^2\phi^2}{f_0^2} - \frac{m^2}{f_0} 
- \frac{2f_0'}{rf_0} \right ) >0
\label{upperbdfn}
\ee
at this point.  An examination of when this combination is never positive
provides an upper bound for $T_c$. (A different $n$ was used in \cite{GKS}, 
which sufficed for the probe limit although it led to a looser bound.
For consistency and comparison between the scalar masses, this
new bound is used here.)

We can also obtain a lower bound by considering $n = 3$. Manipulating 
(\ref{Xeqn}) shows that {\it if} a solution exists, then the integral
\be
\int_{r_+}^\infty \frac{1}{r^3}  \left [\frac{\phi_0^2}{f_0} -m^2 
+ \frac{3f_0}{r^2} - \frac{3f_0'}{r} \right ] = - \int_{r_+}^\infty
\frac{f_0 X_3^{\prime2}}{r^3X_3^2} \leq 0
\label{anlwbd}
\ee
is negative.
Note that negativity of this integral does not imply existence of
a solution to the linearized equation near $T_c$, it is simply a necessary
condition. Since this integral is always negative at large
$T$, and positive as $T\to0$ (for $\kappa^2 \lesssim 0.4$), 
observing where it changes sign provides a lower bound on $T_c$. 
This bound was found to give an extremely reliable indicator 
of $T_c$ as computed numerically.

\subsection{Analytic expansion}

\begin{figure}
\begin{center}
\begin{tabular}{cc}
(a) & (b)\\
\includegraphics[width=7cm]{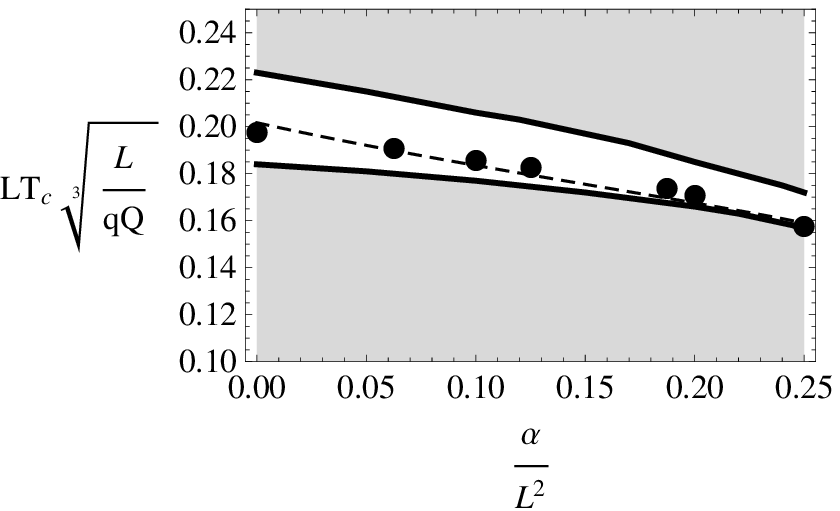} & 
\includegraphics[width=7cm]{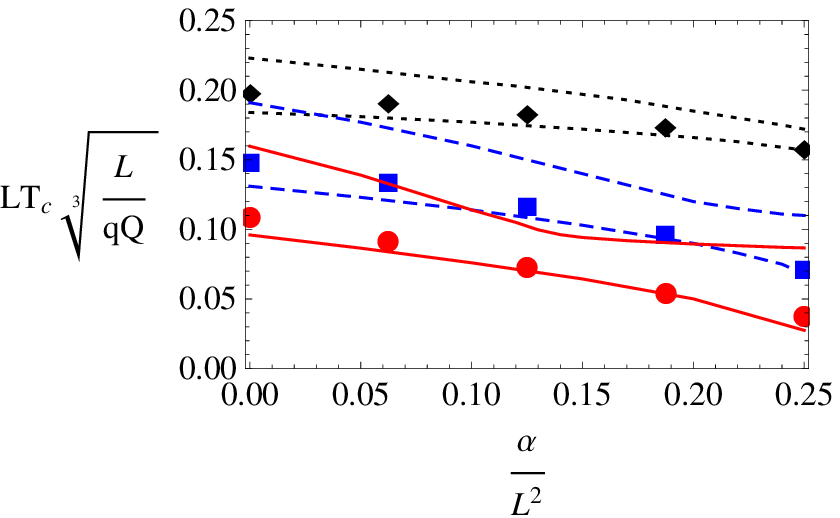} 
\end{tabular}
\end{center}
\caption{
A plot of the critical temperature as a function of $\alpha$ for the
scalar mass $m^2 = -3/L^2$. (a) shows a detailed plot of the probe
limit, with upper and lower bounds shown as solid black lines, the 
matching method value as a dotted line, and numerical data points. 
(b) shows a plot of the upper and lower bounds together with
exact numerical results for the probe limit in black (dotted lines 
and diamond data points), $\kappa^2 = 0.05$ in blue (dashed lines and 
square data points), and $\kappa^2 = 0.1$ in red (solid lines and circular
data points).
}
\label{fig:TcL}
\end{figure}

In instances where the exact solution is only known numerically, it
can be useful to derive an analytic approximation to the exact solution
using a technique of matching. The solution is Taylor expanded around
the horizon and boundary, then matched at a midpoint. This technique 
was used in \cite{GKS} to get an approximation to $T_c$, although the
plot in \cite{GKS} was inaccurate. In brief, we first transform the
radial variable to $z = r_+/r$, and expand the near horizon solution
of a generic field $Y$ as:
\be
Y \simeq Y_0 - Y_1 (1-z) + \textstyle{\frac{1}{2}} Y_2 (1-z)^2
\ee
The equations of motion give:
\bea
\psi_1 &=& \frac{m^2r_+^2}{f_1} \psi_0 \label{psi1def} \\
\psi_2 &=& \frac{\psi_0}{2f_1^2} \left \{ m^2 r_+^2 f_2 
- m^2 r_+^2 f_1 \nu_1 +m^4 r_+^4 - q^2 \phi_1^2 r_+^2 e^{-2\nu_0} 
- 3 m^2 r_+^2 f_1 \right \}\label{psi2def} \\
\phi_2 &=& \phi_1 \left ( 1 + \nu_1 + \frac{2r_+^2q^2\psi_0^2}{f_1} \right)
\label{phi2def} \eea
where $f_1$, $f_2$, and $\nu_1$ are given by (\ref{nueq},\ref{feq}), and
depend on $\phi_0$ and $\psi_0$. Since the asymptotic solutions for
the gauge and scalar field given by (\ref{r:boundary}) also have
two unknowns, matching these fields and their derivatives at some
intermediate point $z_0$ gives four equations which determine these
four unknowns: 
\bea
P - \frac{Qz_0^2}{r_+^2} &=& -\phi_1 (1-z_0) + \frac{\phi_2}{2}
(1-z_0)^2 \label{phict} \\
\frac{-2Qz_0}{r_+^2} &=& \phi_1 - \phi_2 (1-z_0) \label{phipr} \\
\frac{C_+ z_0^{\Delta_+}}{r_+^{\Delta_+}} &=& \psi_0 - \psi_1 (1-z_0)
+ \frac{\psi_2}{2} (1-z_0)^2 \label{psict} \\
\Delta_+ \frac{C_+ z_0^{\Delta_+-1}}{r_+^{\Delta_+}} &=& \psi_1
-\psi_2 (1-z_0) \label{psipr} 
\eea
Clearly, given the complexity of the expressions 
(\ref{psi1def})-(\ref{phi2def}), which themselves contain $f_i$,
this is a rather involved algebraic process in the general 
backreacting and non-critical case. Nonetheless, the system is
tractable analytically if we simply focus on deriving the
critical temperature. In this case, we once again can take the
background forms for $\phi$, $f$ and $\nu$, which considerably
reduce the complexity of the relations: (\ref{phict}) and (\ref{phipr})
are automatically satisfied, $\nu\equiv0$, (\ref{feq}) gives
\bea
f_1 &=& -4 \frac{r_+^2}{L^2} + \frac{\kappa^2 \phi_1^2}{3} \\
f_2 &=& -f_1 + \frac{2\alpha f_1^2}{r_+^2} + 2\kappa^2 \phi_1^2
\eea
and consistency of the $\psi$ equations (\ref{psict}) and (\ref{psipr})
implies
\be
2\Delta_+ \psi_0 - 2 \left ( \Delta_+ + (1-\Delta_+)z_0 \right ) \psi_1
+ \left ( \Delta_+ - 2(\Delta_+-1) z_0 + (\Delta_+ - 2)z_0^2 \right) \psi_2
=0
\ee
Although this is a cumbersome expression for nonzero $\kappa^2$, 
it is just a quadratic for $\phi_1^2/r_+^2=4Q^2/r_+^6$, and a linear
relation in the probe limit, which in either case can be 
straightforwardly solved and the critical temperature obtained from
\be
T_c = r_+ \left [ \frac{1}{\pi L^2} - \frac{\kappa^2 Q^2}{3 r_+^6} \right]
\ee
It only remains to choose a value for $z_0$. In \cite{GKS} the value 
$z_0 = 1/2$ was chosen to be specific, however, this choice does not
take into account the variation of the effective adS length scale with
$\alpha$. In essence, the matching point should be chosen to
represent where the asymptotic expansion $f \simeq r_+^2/z^2L_e^2$
turns over into the near horizon expansion. This obviously depends
on the relative ratio of $r_+$ to $L_e$, although as $\alpha\to L^2/4$
this point moves rather significantly towards the boundary. 
A good approximation to the departure from the asymptotic regime is
$z_0 \simeq \alpha/L^2 + 1/2$, which is the value used in figure
\ref{fig:TcL}(a).
It must be noted however, that there is a degree of arbitrariness in
this matching process. Although the choice of $z_0$ attempts to take 
into account the likely variation of the domains of validity of each 
expansion with $\alpha$,
this is a rather ad hoc process, and as backreaction is switched on,
the matching procedure becomes less and less reliable. It is included here
for completeness, however by far the best analytic guide to
critical temperature is the lower bound.

\subsection{Discussion of analytic results}

\begin{figure}
\begin{center}
\includegraphics[width=10cm]{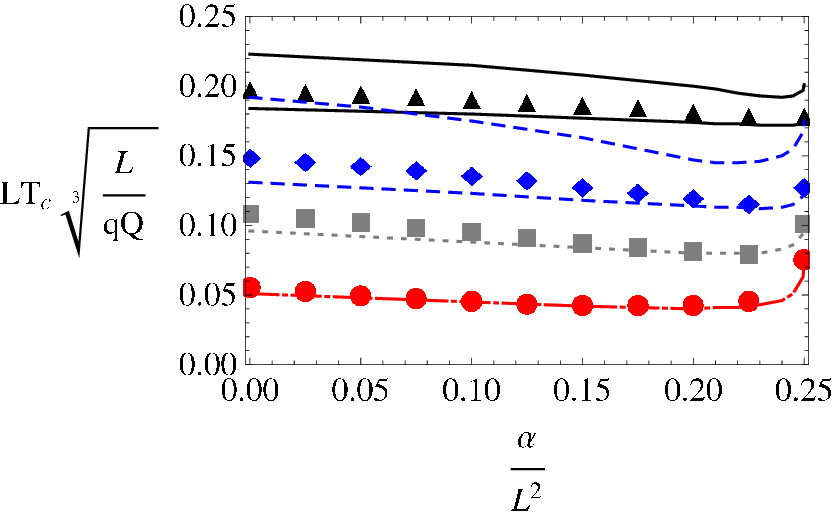}
\end{center}
\caption{
A plot of the critical temperature as a function of $\alpha$ for a
selection of $\kappa^2$. The analytic bounds are shown as lines and
the numerical data as points. Respectively: $\kappa^2=0$ is shown in
black, with solid lines and triangular data points; $\kappa^2 = 0.05$
has blue dashed lines and diamonds; $\kappa^2=0.1$ has a grey dotted 
line and squares; $\kappa^2=0.2$ has a red dot-dash line with circular 
data points. The lower bound is shown for all $\kappa^2$ values, but 
the upper bound is shown only for the lowest two values of $\kappa^2$, 
as they overlap significantly with the other data and confuse the plot.
}
\label{fig:anbound}
\end{figure}

By examining the behaviour of a putative condensed scalar in the
bulk, it is possible to derive upper and lower bounds to the critical 
temperature. A matching method can also be used to get an approximate
solution for the scalar, however, this method (as well as the upper bound)
becomes unreliable as the backreaction is increased. On the other
hand, the lower bound appears to be extremely representative of the 
behaviour of the actual critical temperature, and works well for
a significant range of backreaction. Figures \ref{fig:TcL} 
and \ref{fig:anbound} show the analytical results, and will be
discussed further together with numerical data in the next section.

\section{Backreacting superconductors}
\label{sec:BR}

In order to find the actual behaviour of the bulk superconductor,
in \cite{GKS} and \cite{BGKS} we integrated (\ref{phieq}$-$\ref{feq}) 
numerically. As already stated, we took $L=Q=q=1$, and varied $r_+$ 
to study how the system reacted to varying temperature.
From these numerical solutions, we can deduce the
critical temperature, determine the behaviour of the condensate
with temperature, and compute the conductivity of the holographic
superconductor as discussed in subsection \ref{sec:cond}.

\subsection{The dependence of $T_c$ on $\alpha$ and $\kappa^2$.}
\label{sec:Tc}

The analytic bounds for $T_c$ have been plotted in figures \ref{fig:TcL}
and \ref{fig:anbound} together with the exact values of $T_c$ obtained 
by numerical computation for both masses of the scalar field. For
$m^2 = -3/L^2$, both bounds are shown, as well as the matching method
result for the probe limit. For $m^2 = -3/L_e^2$,
the upper bound has only been shown for $\kappa^2 \leq 0.05$, as above this
value it becomes less predictive and clutters the plot, and indeed
beyond $\kappa^2\sim 0.2$ (corresponding to $q \sim 2.25$ 
in the notation of \cite{HHH}) it ceases to have quantitative
value for any $\alpha$. The lower bound
on the other hand becomes successively more accurate as the values of
$\kappa^2$ are stepped up, and gives a very good quantitative guide 
to the behaviour of $T_c$ as we vary $\alpha$ and $\kappa^2$.

It is easy to see that in all cases, the effect of backreaction is to
decrease $T_c$ and thus make condensation harder.  
We can see this by using Gubser's rough argument, \cite{Gub}, that
the effective scalar mass $m_{\rm eff}^2=m^2-q^2\phi(r)^2/f(r)$
becomes more negative as backreaction is turned up.
Essentially, the effect of backreaction is that the condensation of the scalar
field not only screens the charge of the black hole, but also its mass,
as the scalar and gauge fields now contribute to the ADM mass. This means
that for a given charge and temperature, the radius of the black hole
is increased, which makes it harder for the scalar to condense.

One very interesting feature clearly exhibited in the bounds is the 
turning point in $T_c$ as a function of $\alpha$ for the mass $m^2 = -3/L_e^2$.
In the probe case, this occurs very near the Chern-Simons
limit $\alpha = L^2/4$, and is barely perceptible in the numerical data, 
however 
once backreaction is switched on, the minimum becomes much more pronounced,
and indeed for large backreaction ($\kappa^2 = 0.2$) the Chern-Simons
limit is showing a considerable enhancement of $T_c$ over the typical 
values for lower $\alpha$.

\subsection{The scalar condensate.}
\label{sec:O3}

\begin{figure}
\includegraphics[width=7.2cm]{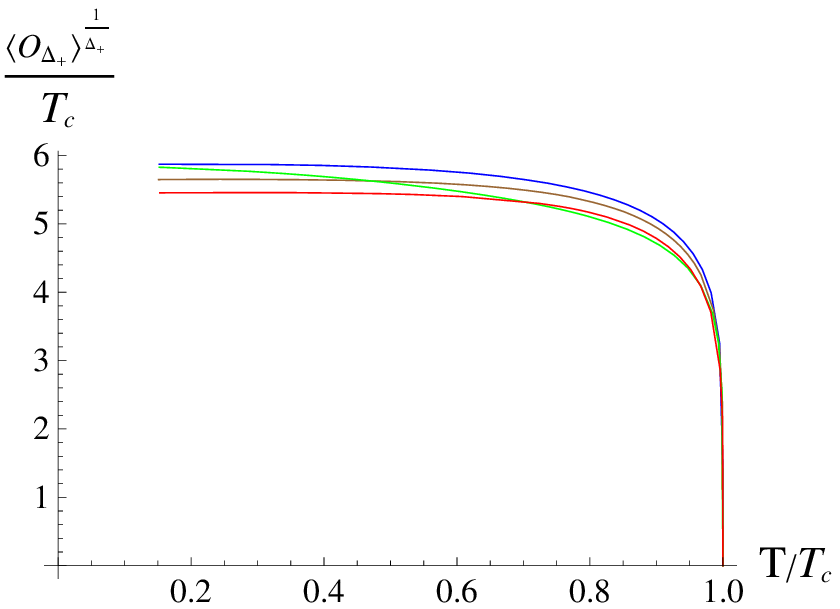} 
\vskip -5cm \hskip 8cm
\includegraphics[width=5.2cm, angle=270]{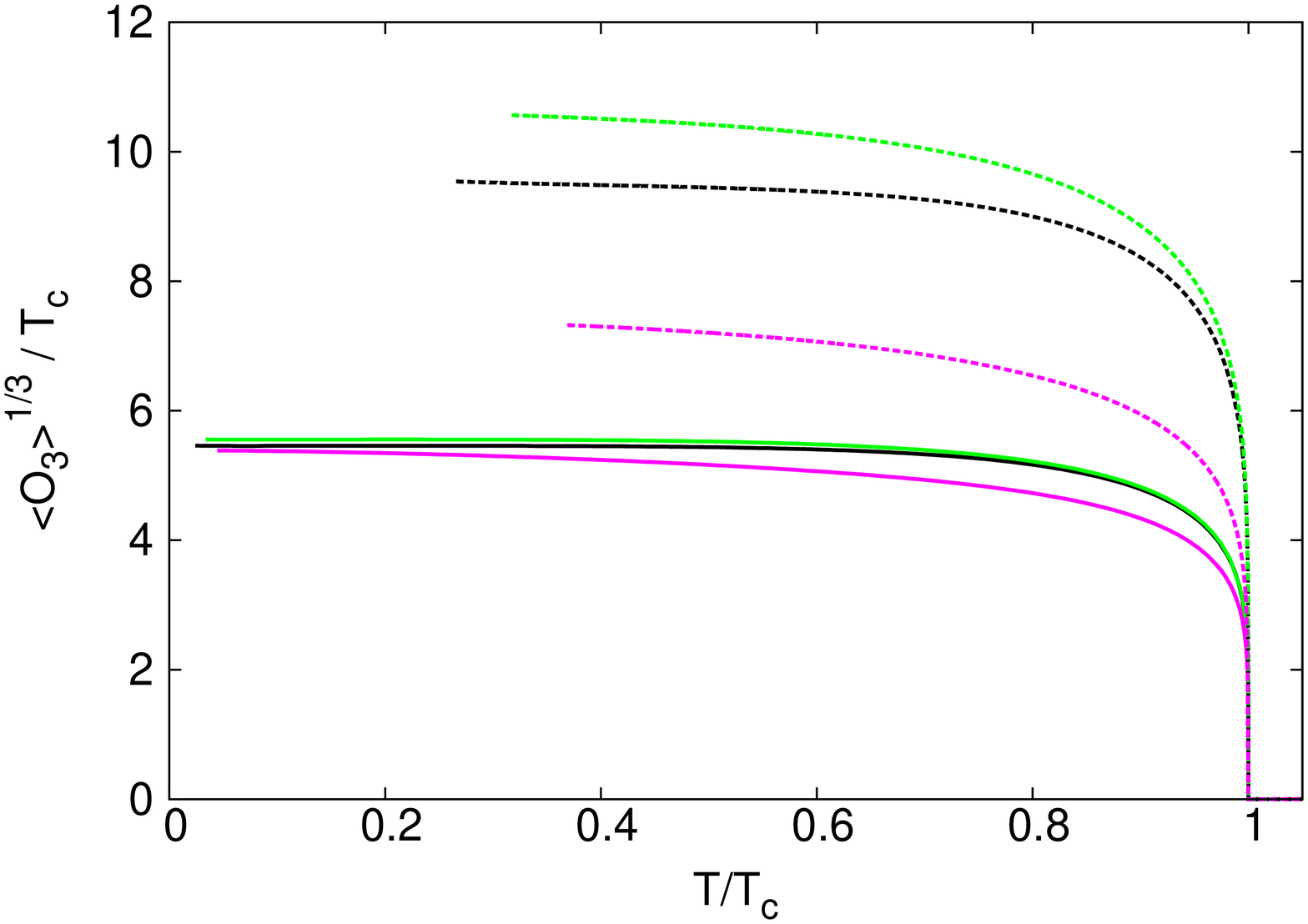} 
\caption{Two plots of the condensate as a function of temperature for
$m^2 = -3/L^2$ (left) and $m^2 = -3/L_e^2$ (right). The plot on the left
shows several values of $\alpha$ in the probe limit: $\alpha = 0$ in red,
the lowest line, then $\alpha = 0.1$ in brown, $0.2$ in blue and $0.25$
in green. The plot on the right shows $\kappa^2=0$ in solid lines,
and $\kappa^2 = 0.1$ as dotted lines.
The black plot is $\alpha=0$, green is $\alpha=0.125$ and magenta
is $\alpha=0.25$.
}
\label{fig:TcPlot}
\end{figure}

The numerical results allow a full exploration of the dependence of the 
condensate on temperature, and we expect to see the condensate switching
on sharply at the critical temperature, then saturating at low temperatures,
and this is indeed what is seen. Figure \ref{fig:TcPlot} 
shows $\langle {\cal O}_{\Delta_+} \rangle^{1/{\Delta_+}}$
as a function of temperature for both scalar masses.

Each line in the plot shows the characteristic curve of the condensate
$\langle {\cal O}_{\Delta_+} \rangle^{1/{\Delta_+}}$ condensing at 
some critical temperature. For $m^2 = -3/L^2$ the curves are shown in 
the probe limit. While the critical temperature does not display the 
minimum in $\alpha$, it can nonetheless be seen that the Chern-Simons 
limit does look rather different, and the scalar field saturates far
more slowly as the temperature drops. For $m^2 = -3/L_e^2$, three different
values of $\alpha$ ($0, 0.125, 0.25$) and two different values 
of $\kappa^2$ ($0, 0.1$) are chosen to display the features of 
the system.  
Both plots are shown with the curves normalized by $T_c$.  In the latter 
plot, the effect of $\kappa^2$ is to increase the height of these graphs,
in spite of the fact that the raw data tends to have lower values
of $\langle {\cal O}_3 \rangle$. This is clearly because the most
significant impact of increasing gravitational backreaction is
that the critical temperature of the system is lowered.
In this case the effect of backreaction is
extremely marked, with the condensate varying more widely with $\alpha$.

\subsection{Conductivity}
\label{sec:cond}

In \cite{HorRob}, Horowitz and Roberts observed an interesting 
phenomenon for the conductivity of the boundary theory.  They 
considered the model (\ref{action}) in the probe Einstein limit for a range
of different bulk scalar masses, and on computing the conductivity 
found an apparent universal relation
\begin{eqnarray}
\frac{\omega_g}{T_c} \simeq 8 \ ,
\end{eqnarray}
with deviations of less than 8 \%. 

Conductivity is conventionally expressed as the current density
response to an applied electric field:
\be
\sigma = \frac{\cal J}{\cal E} \, .
\label{condclass}
\ee
The boundary four-current $J_\mu$ is dual to the bulk field $A_\mu$,
thus we must consider perturbations of $A_\mu$ to compute the conductivity.
The full perturbation problem must take into account perturbations
of the metric ($h_{ti}$ and $h_{ri}$) and the perturbation of the
Gauss-Bonnet tensor $H_{ti}, H_{ri}$. After some algebra, these
reduce to:
\begin{eqnarray}
{\dot h}^\prime_{ti}-\frac{2}{r}{\dot h}_{ti}-{\ddot h}_{ri}
+\frac{L^2fe^{2\nu}}{r^2 - 2\alpha f}
\left( 1 - \frac{\alpha(2\nu'f + f')}{r} 
\right) \Delta h_{ri}
+\frac{2\kappa^2r^2 {\dot A}_i\phi^\prime}{r^2-2\alpha f} 
&=&0 \;\;\;
\label{ri}\\
\frac{e^{-\nu}}{rf} \left [ r f e^\nu A_i' \right ]'
-\frac{{\ddot A}_i}{f^2e^{2\nu}}
+\frac{L^2}{r^2f}\Delta A_i
-\frac{2}{f}q^2\psi^2A_i
+\frac{\phi'}{fe^{2\nu}}
\left( h_{ti}^\prime-\frac{2}{r}h_{ti} - {\dot h}_{ri} \right)
&=&0 \label{A1}
\end{eqnarray}
where $h_{ab}$ is the perturbation of the metric tensor, and
$A_i$ is the perturbation of the gauge field, which has only
spatial components.
Writing $A_i(t,r,x^i)=A(r)e^{i{\bf k}\cdot{\bf x}-i\omega t}e_i$, 
and setting ${\bf k}={\bf 0}$, (\ref{ri}) can be integrated and substituted
in (\ref{A1}) to obtain:
\be
A^{\prime\prime}+\left(\frac{f^\prime}{f}+\nu^\prime+\frac{1}{r}\right)A^\prime
+\left[\frac{\omega^2}{f^2e^{2\nu}}-\frac{2}{f}q^2\psi^2
-\frac{2\kappa^2r^2\phi^{\prime2}}{fe^{2\nu}\left(r^2-2\alpha f\right)}
\right]A =0 \;.
\label{A:eq}
\ee

This is solved under the physically imposed boundary condition of no
outgoing radiation at the horizon:
\be
A(r) \sim f(r)^{-i\frac{\omega}{4\pi T_+}} \ ,
\label{ingoing}
\ee
where $T_+$ is the temperature.
In the asymptotic adS region $(r\rightarrow\infty)$, 
the general solution takes the form
\be 
\label{genasoln}
A=a_0 + \frac{a_2}{r^2}
+\frac{a_0 L_e^4 \omega^2}{2r^2}
\log \frac{r}{L}
\ee
where $a_0$ and $a_2$ are integration constants. 
Note there is an arbitrariness of scale in the logarithmic term,
as pointed out in \cite{HorRob}, however, this is related
to an arbitrariness in the holographic renormalization process;
see Appendix A of \cite{BGKS} for a full computation of the conductivity
and discussion of this renormalization scale. After careful computation 
using the method of Skenderis \cite{Skenderis}, the conductivity of 
the EGB system is found to be \cite{BGKS}:
\be\label{ConductivityEquation}
\sigma=\frac{2a_2}{i\omega L_e^4 a_0 } 
+\frac{i\omega}{2} - i\omega \log \left ( \frac{L_e}{L} \right) \ .
\ee
Note that the imaginary term linear in $\omega$ has an arbitrariness
of scale from the counterterm subtraction, and this is capitalized
in the presentation of numerical data, where a suitable 
renormalization scale is chosen to make the features of the plot clearest.
\begin{figure}
\begin{center}
\begin{tabular}{cc}
\includegraphics[width=5.2cm]{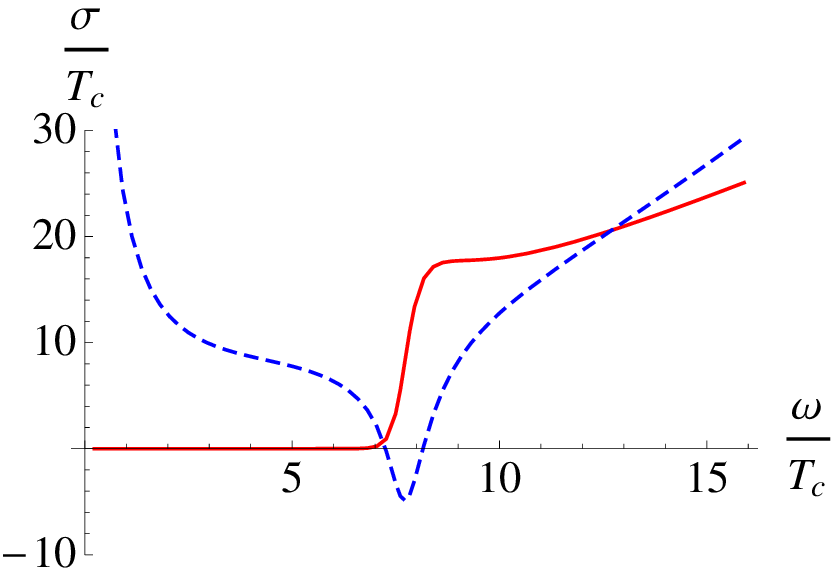} &
\includegraphics[width=5.2cm]{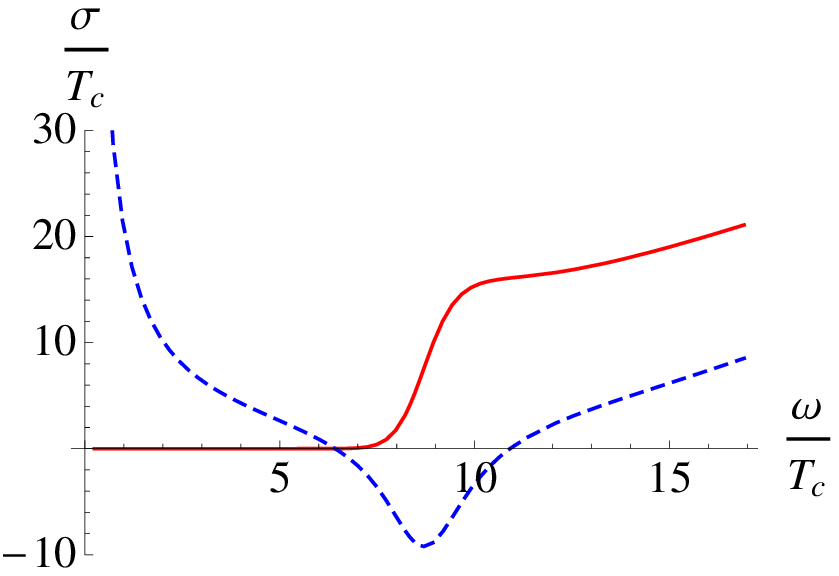} \\(a) & (b)\\
\includegraphics[width=5.2cm]{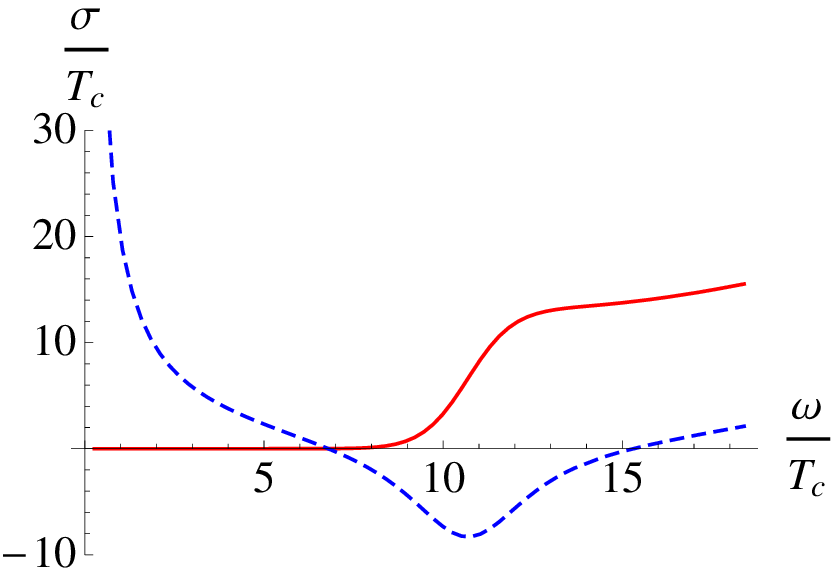} & 
\includegraphics[width=5.2cm]{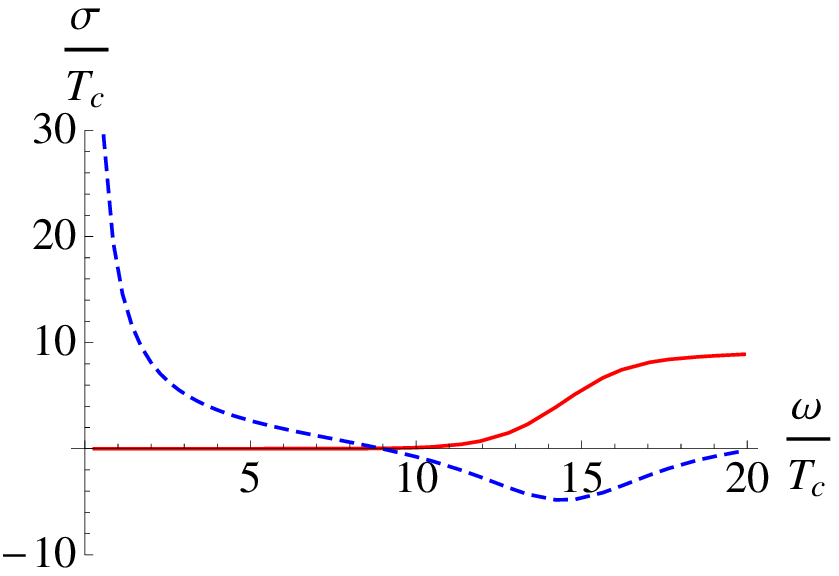} \\
(c) & (d)
\end{tabular}
\end{center}
\caption{Conductivity: a range of plots showing the real (solid red line)
and imaginary (blue dashed line) parts of the (normalized) conductivity 
as a function of (normalized) frequency for the scalar mass $m^2 = -3/L^2$. 
Plot (a) is $\alpha = 0$, (b) is $\alpha = 0.1$, and (c) is $\alpha=0.2$,
and (d) is $\alpha = 0.25$.
}
\label{fig:ConductL}
\end{figure}

\begin{figure}
\begin{center}
\begin{tabular}{cc}
\includegraphics[width=5.2cm, angle=270]{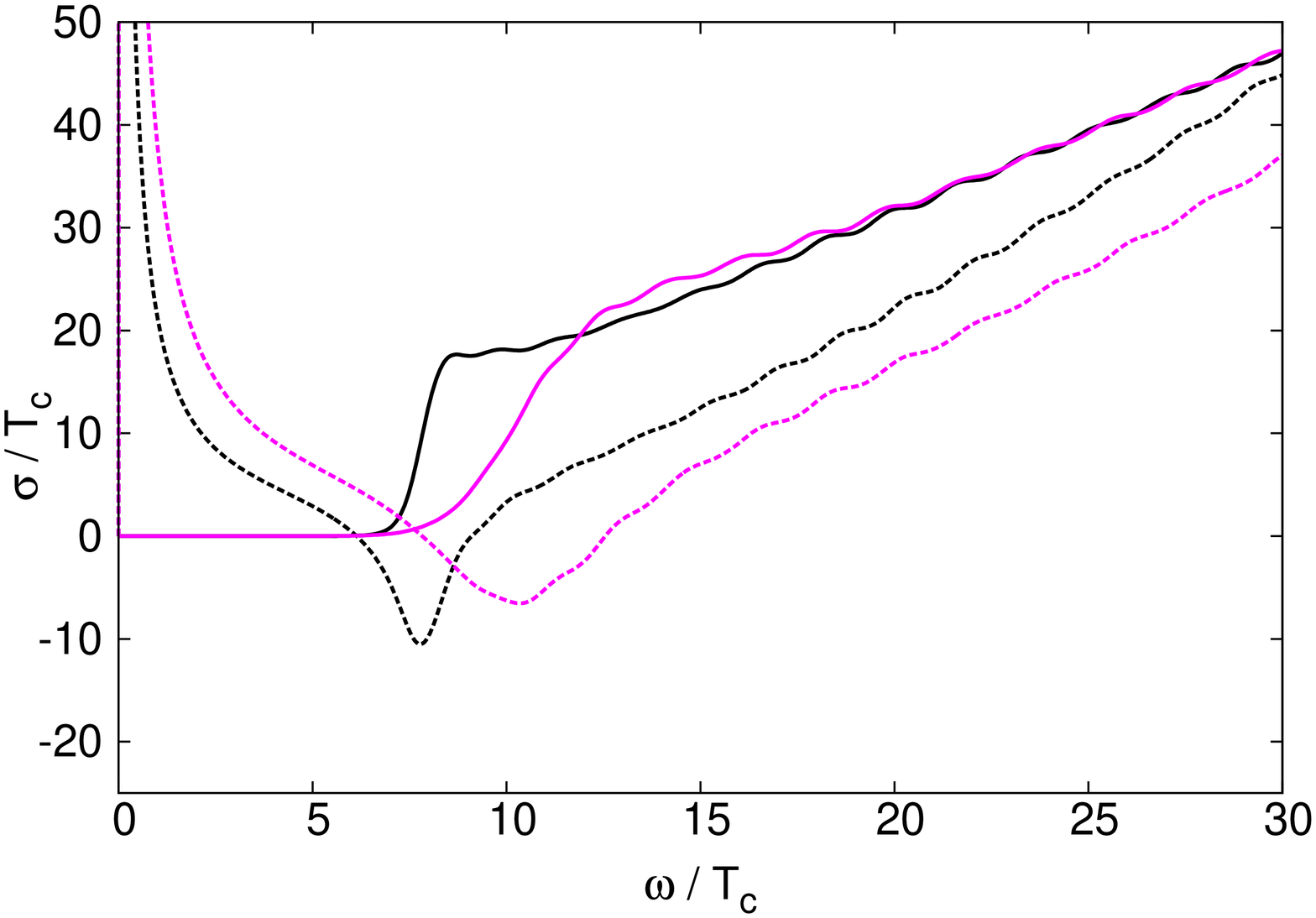} &
\includegraphics[width=5.2cm, angle=270]{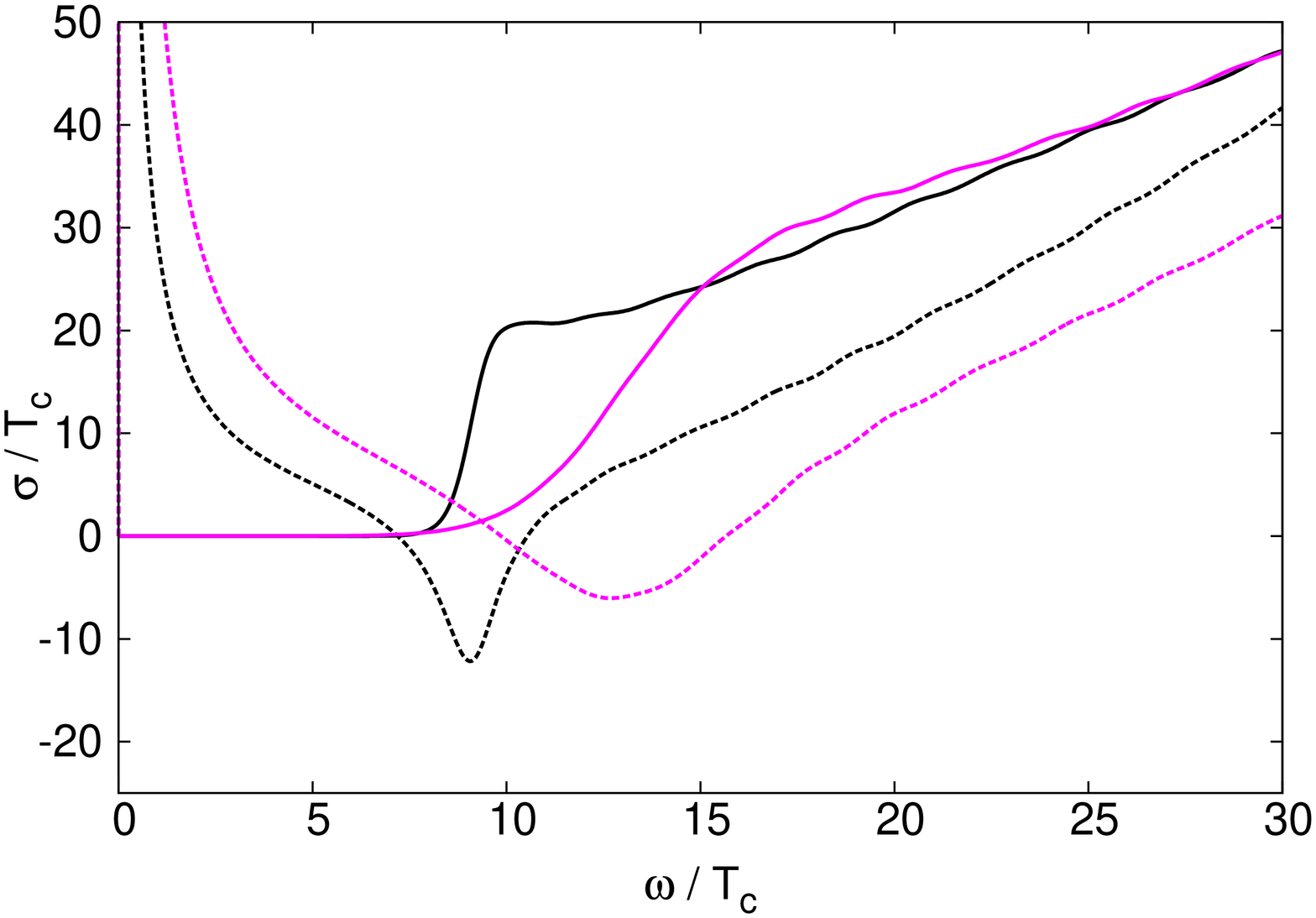} \\
(a) & (b) 
\end{tabular}
\end{center}
\includegraphics[width=5.2cm, angle=270]{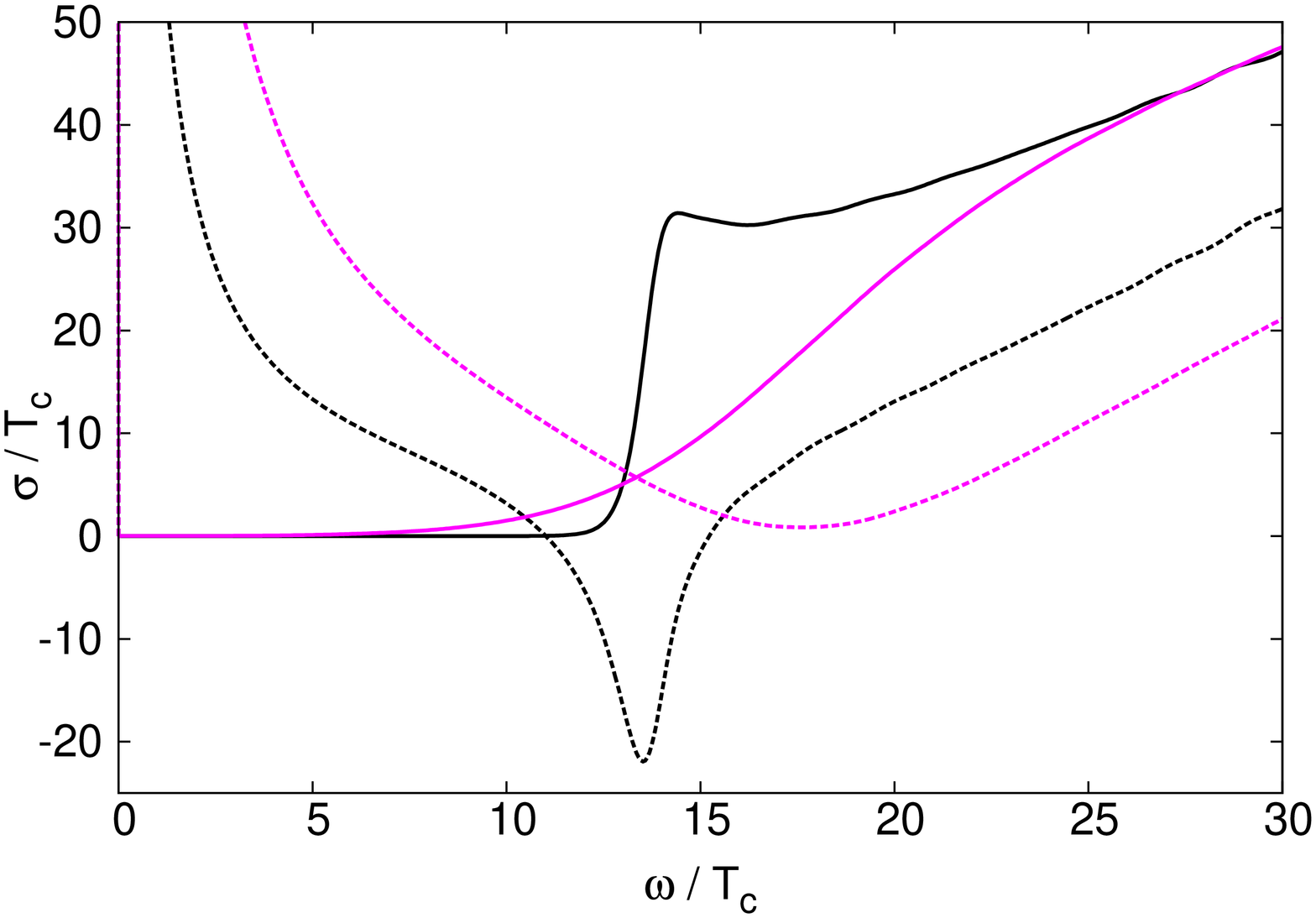}
\begin{center}
\vskip -5cm \hskip 8cm
\includegraphics[width=7cm]{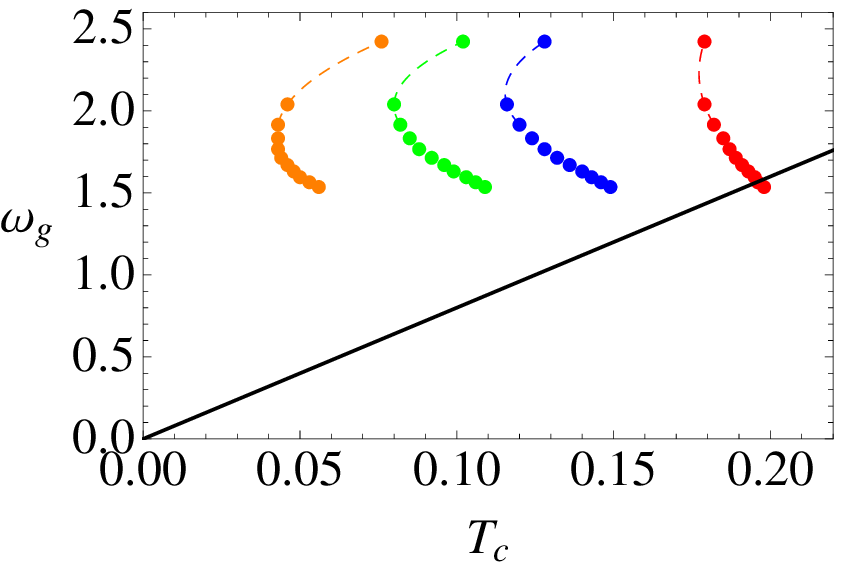}
\begin{tabular}{cc}
(c) \;\;\;\;\;\;\;\;\;\;\;\;\;\;\;\;\;\;\;\;\;\;\;\;\;\;\;\;\;\;\;\;\;\;\;\;
& \;\;\;\;\;\;\;\;\;\;\;\;\;\;\;\;\;\;\;\;\;\;\;\;\;\;\;\;\;\;\;\;\;\;\;\; (d)
\end{tabular}
\end{center}
\caption{Conductivity: a range of plots showing the conductivity
of the $m^2 = -3/L_e^2$ superconductor. Plots (a) $-$ (c) show the
real (solid line) and imaginary (dashed line) parts of 
the conductivity as a function of frequency for 
$\alpha = 0, 0.125$, and $0.25$ respectively.
In each case the conductivity is shown for no backreaction
in black, and for a backreacting parameter $\kappa^2 = 0.05$ in magenta.
The slight undulations in the plots at large $\omega$ is a numerical
artefact.
Plot (d) shows the gap frequency as a function of $T_c$ with the line
$\omega_g=8T_c$ is shown in black. The different colours 
represent from right to left: 
Red is $\kappa^2 = 0$, Blue is $\kappa^2 = 0.05$, Green $0.1$, and
Orange $\kappa^2 = 0.2$. In each case $\alpha$ is incremented
from $0$ to $0.25$. As the gap alters rapidly near the Chern-Simons
limit, the dotted lines are added by hand to guide the eye.
}
\label{fig:Conduct}
\end{figure}

Figure \ref{fig:ConductL} shows the real and imaginary parts of 
the conductivity as a function of $\omega/T_c$, calculated
for the scalar mass $m^2 = -3/L^2$ in the probe limit and for a 
selection of values of $\alpha/L^2$. These plots clearly show
that as $\alpha$ increases, the normalised gap frequency shifts 
to higher values. An additional feature is that the gap becomes 
softer with increasing $\alpha$.
These plots should be contrasted with figure \ref{fig:Conduct},
which shows similar conductivity plots, but now for scalar mass
$m^2 = -3/L_e^2$, and with both no backreaction and a 
backreaction of $\kappa^2 = 0.05$. Three sample values 
of $\alpha$ are shown: $0$, $0.125$, and $0.25$.
In these plots, the gap if anything gets harder with increasing
$\alpha$, however once backreaction is included, the gap 
becomes more gentle and extended, and while the dip in 
$\textsc{Im}(\sigma)$ is smoothed, it is still clearly apparent. 

In all of these plots, the gap is clearly indicated by a rise in the 
real part of $\sigma$, which coincides with
the global minimum of $\textsc{Im}(\sigma)$.  As already noted, 
the imaginary part of (\ref{ConductivityEquation})
is only valid up to a linear term in $\omega$, the size of which is 
dependent on the renormalization scheme employed (and also on the 
charge $Q$). One can therefore tune this linear term in
$\textsc{Im}(\sigma)$ to create a finite global minimum 
if it is not initially present, and indeed the plots in figure
\ref{fig:ConductL} differ from those presented in \cite{GKS} 
by precisely this sort of term so as to make the minimum apparent.
This minimum can therefore be used to define $\omega_g$, the value 
of the frequency gap. 

The frequency gap is a distinct characteristic of a superconductor and in   
the BCS theory of superconductivity this frequency gap corresponds to the 
minimum energy required to break a Cooper pair.  As mentioned above, 
in \cite{HorRob} it was claimed that for the holographic superconductor 
the relation $\omega_g/T_c\simeq8$ had a certain universality, proving 
stable for a range of scalar masses and dimensions.  In both
\cite{GKS} and \cite{BGKS}, this
relation was shown to be unstable to Gauss-Bonnet corrections.
Figure \ref{fig:Conduct}(d) in particular gives a very clear indicator 
of how backreaction and higher curvature terms affect the gap.
Increasing either $\alpha$ or $\kappa^2$ increases $\omega_g/T_c$.
For the case of increasing $\alpha$, the effect occurs mainly because
of a shift in the gap, rather than a significant alteration
of $T_c$, which varies much more strongly with backreaction than
$\alpha$. On the other hand, varying $\kappa^2$ practically does not
alter $\omega_g$ at all, whereas $T_c$ drops dramatically, leading to 
a sharp rise in $\omega_g/T_c$.

\section{Summary}
\label{sec:disc}

This presentation reviewed the work of \cite{GKS} and \cite{BGKS}
on exploring the implications of Gauss-Bonnet corrections to
holographic superconductors.  
The results show that increasing backreaction
lowers the critical temperature of the superconductor hence 
increasing $\omega_g/T_c$. The effect of higher curvature terms
is more subtle. Although these initially act in a similar fashion
to backreaction in lowering the critical temperature, for
significant GB coupling and larger scalar masses,
the critical temperature eventually begins
to increase. The conductivity gap is also modified, with both 
$\omega_g$ and $T_c$ altering to increase the ratio $\omega_g/T_c$.
Clearly higher dimensional holographic superconductors have a rich
structure, with or without higher curvature 
corrections.

\ack
I would like to thank Luke Barclay, Sugumi Kanno, 
Jiro Soda and Paul Sutcliffe for being such lively and pleasant
collaborators. I would also like to thank Christos Charmousis,
Rob Myers and Simon Ross for useful conversations. I acknowledge
the support of STFC under the rolling grant ST/G000433/1.

\end{document}